\documentclass[twocolumn,superscriptaddress,preprintnumbers,amsmath,amssymb]{revtex4}
\usepackage{graphicx}
\usepackage{dcolumn}
\usepackage{verbatim}
\usepackage{bm}

\begin{document}
\title{Ground state fidelity and quantum phase transitions in free Fermi systems} 
\author{Paolo Zanardi}
\affiliation{Institute for Scientific Interchange, Villa Gualino, Viale
Settimio Severo 65, I-10133 Torino, Italy}
\author{Marco Cozzini}
\affiliation{Institute for Scientific Interchange, Villa Gualino, Viale
Settimio Severo 65, I-10133 Torino, Italy}
\affiliation{Dipartimento di Fisica, Politecnico di Torino, Corso Duca degli
Abruzzi 24, I-10129 Torino, Italy}
\author{Paolo Giorda}
\affiliation{Institute for Scientific Interchange, Villa Gualino, Viale
Settimio Severo 65, I-10133 Torino, Italy}

\date{\today}

\begin{abstract}
We compute the fidelity between the ground states of general quadratic
fermionic hamiltonians and analyze its connections with quantum phase
transitions. Each of these systems is characterized by a $L\times L$ real
matrix whose polar decomposition, into a non-negative $\Lambda$ and a unitary
$T$, contains all the relevant ground state (GS) information. The boundaries
between different regions in the GS phase diagram are given by the points of,
possibly asymptotic, singularity of $\Lambda$. This latter in turn implies a
critical drop of the fidelity function. We present general results as well as
their exemplification by a model of fermions on a  totally connected graph.

\end{abstract}
\maketitle

{\em Introduction}--
Quantum matter at zero temperature is known to exist  in different phases that
cannot be continuously deformed into each other without giving rise to a singular
behaviour of some physical quantity. This singularity, which extends its
influence over a finite range of temperature, is due to the competition
between  different kinds of quantum fluctuations trying  to order the ground
state  according to distinct and alternative correlation patterns. These
phenomena, known as quantum phase transitions (QPTs) recently attracted a big
deal of attention \cite{sachdev}. In particular ideas  drawn from Quantum
Information \cite{qis} e.g., quantum entanglement, have been proven to be
conceptually insightful as well as to provide effective tools to investigate
QPTs \cite{qpt-qis}.

In this paper we aim at further investigating the approach to QPTs advocated in
Ref.~\cite{za-pa}.
There it was suggested, inspired by  the results of Ref.~\cite{zanardi-china}, that the boundaries between different quantum phases
can be analyzed in terms of an extremly simple quantity: the overlap i.e., the scalar product, between the ground states (GSs) 
corresponding 
to slightly different values of the coupling constants. The critical points being characterized by a large, possibly discontinuous, drop of the overlap.
More interestingly, in Ref.~\cite{za-pa} it has been shown that, for some
models, the overlap contains information about finite-size scaling and universality
classes.
We shall use a quantity strictly related to the overlap,
the {\em fidelity} \cite{qis,zhou-barja}.

We shall deal with 
general systems of free-fermions \cite{stephan}.
This is a class of quantum models that are i) physically relevant ii) simple enough to allow for a rather comprehensive
analytical treatment and, at the same time iii) sufficiently rich to feature non-trivial GS phase diagrams.
More specifically the class of fermionic systems we are going to consider is the one described by the following quadratic Hamiltonians
\begin{equation}
H=\sum_{i,j=1}^L c_i^\dagger A_{i,j} c_j +
1/2 \sum_{i,j=1}^L(c_i^\dagger B_{i,j} c_j^\dagger +\rm{h.c.}) \ ,
\label{ham}
\end{equation}
where: the $c_i$'s ($c_i^\dagger$'s) are the annihilation (creation) operators
of $L$ fermionic modes, $A,B\in M_L(\mathbb{R})$
are $L\times L$ {\em real}
matrices, symmetric and anti-symmetric respectively i.e., $A^T=A,\, B^T=-B$. 

The Hamiltonian family (\ref{ham}), and their ground states, can be directly parametrized by the generic {\em real} matrix $Z$
and the defining $A:=(Z+Z^T)/2$ and $B:=(Z^T-Z)/2.$ From this perspective the space of coupling constants of (\ref{ham})
is the $L^2$-dimensional full matrix algebra $M_L(\mathbb{R})$ even though 
 most of the $Z$'s
will give raise to rather unphysical Hamiltonians e.g., highly non-local. 
In order to obtain physically meaningful models one might take, for example,
$A$ to be the adiacency matrix of some graph whose $L$ vertices label the
fermionic modes.

The central quantity of our study is the GS fidelity \cite{losch}
\begin{equation}
{\cal F}(Z,\tilde{Z}):=|\langle\Psi_{Z},\, \Psi_{\tilde{Z}}\rangle| \ ,
\label{over}
\end{equation}
where $|\Psi_Z\rangle$ denotes the GS corresponding to the Hamiltonian (\ref{ham}).
In the case in which $\tilde{Z}=Z+\delta Z$ we will write ${\cal F}(Z,\tilde{Z})=e^{-S(Z)}.$
The whole idea of the present approach is that QPTs are 
characterized by a singular behaviour
i.e., drop (enhancement), 
of ${\cal F}(Z)$ ($S(Z)$).

{\em The Ground State}--
Let us start by an explicit analytical characterization of the ground state of the quadratic fermionic Hamiltonian
(\ref{ham}). 
To this purpose we use the fact that,
for  even number of particles and sites, the (unnormalized) ground state of (\ref{ham}) is given by \cite{peschel}
\begin{equation}
|{\tilde\Psi}_{Z}\rangle=\exp\left(1/2 \sum_{i,j=1}^L c_i^\dagger G_{i,j} c_j^\dagger\right )|0\rangle \ ,
\label{Psi}
\end{equation}
where $c_i|0\rangle=0, \forall i$ and 
$G$ is a $L\times L$ anti-symmetric matrix, determined by $g G+h=0$
where $g=1/2(\Phi+\Psi),\, h=1/2(\Phi-\Psi) \in M_L(\mathbb{R})$ and
$\Phi(A-B)(A+B)=\Lambda^2 \Phi, \Lambda \Psi=\Phi(A-B),\, \Lambda={\rm{diag}}(\Lambda_1,\ldots,\Lambda_L)\ge 0$
\cite{li-ma}.
The  $\Lambda_i$'s comprise the {\em single-particle} energy spectrum.
From these equations, by recalling $Z=A-B$ and by assuming $\Lambda$ (and hence $Z$) to be {\em invertible} 
 one can immediately obtain
$g = \Phi (\openone +\Lambda_\Phi^{-1} Z )/2$ and 
$h = \Phi (\openone-\Lambda_\Phi^{-1} Z)/2$
where $\Lambda_\Phi:= \Phi^{-1}\Lambda\Phi.$
From which, {\em if} $g$ is invertible, it follows
\begin{equation}
G=\frac{\Lambda_\Phi^{-1}Z-\openone}{\Lambda_\Phi^{-1}Z+\openone}=\frac{T-1}{T+1}=\frac{T^{1/2}-T^{-1/2}}{T^{1/2}+T^{-1/2}} \ ,
\label{G}
\end{equation}
where $T:=\Lambda_\Phi^{-1}Z.$
The eigenvalue equation above can be rewritten as
$Z Z^\dagger= \Lambda_\Phi^2,$ from which it follows that 
$T$ is the unitary part of the polar decomposition of $Z=\Lambda_\Phi T$ \cite{bathia}.
The operator $g$ is not invertible \textit{iff} $-1\in\rm{Sp}(T),$ let be $p$ its degeneracy. 
If $p$ is even (odd) then it is easy to find out a canonical transformation,
parity-conserving (parity-flipping), mapping $T$ into a $T^\prime$ where all
the $-1$'s  are transformed into $1$'s \cite{cozzini}. 
The corresponding $g^\prime$ becomes then invertible and one has a GS with the form (\ref{Psi}). This also implies
that $\det{T}=(-1)^p$ corresponds to the GS  in the  sector with parity $(-1)^{\sum_j n_j}=(-1)^p.$
Since the parity  is conserved by the Hamiltonian (\ref{ham}), its change in the GSs is always related 
to level-crossings i.e., {\em first order} quantum phase transitions, 
and these of course give rise to a discontinuous drop to zero of the fidelity. 
Now we are in the position of proving the following 

{\em Proposition 0}
If the  ground state of (\ref{ham}) is in the even-number of particles sector then it has the form
\begin{equation}
|\Psi_Z\rangle=\otimes_{\nu=1}^{L/2} [\cos(\theta_\nu/2) |00\rangle_{\nu,-\nu} + \sin (\theta_\nu/2)  |11\rangle_{\nu,-\nu}] \ ,
\label{Psi_D}
\end{equation}
where $\{e^{\pm i\theta_\nu}\}_{\nu=1}^{L/2}={\rm{Sp}}(T)$ and 
 $|0\rangle_\nu$  ($|1\rangle_\nu$) denotes the vacuum (occupied) state of the new pairs of fermionic modes $\tilde{c}_{\pm \nu}$ obtained
from the $c_j$'s by means of a suitable $L\times L$ unitary $U.$

{\em Proof.}
From Ref. \cite{schl} it follows that $G$ can be brought by an unitary
transformation $U$ into the canonical block form $G=U({\bf{0}}_{L-2M}\oplus
G_D)U^T$,
$G_D=
i \oplus_{\nu=1}^{M} t_\nu \sigma_{(\nu)}^y 
$.
Here with $\sigma_{(\nu)}^y$ we denote a $\sigma^y$ Pauli matrix  acting on the two-dimensional span of two single-particle modes
(conventionally labelled by $\nu$ and $-\nu$) and $t_\nu\in\mathbb{R}-\{0\}.$
By allowing the first $L-2M$ (an even number) $t_\nu$'s to be zero we can write $G=U G_D U^T$ where the sum over $\nu$ goes now from one to $L/2$.
By redefining the fermionic modes via $U$ and normalizing the state vector (\ref{Psi}) $|\Psi_Z\rangle:= |\tilde{\Psi}_Z\rangle/\|\tilde{\Psi}_Z\rangle\|$ one then gets
$ |\Psi_Z\rangle=\otimes_{\nu=1}^{L/2} [c_\nu |00\rangle_{\nu,-\nu} +s_\nu  |11\rangle_{\nu,-\nu}]$,
 with $c_\nu:=(1+t_\nu^2)^{-1/2}, s_\nu:= t_\nu (1+t_\nu^2)^{-1/2}$. 
From the  block decomposition of $G$ \cite{schl}  it follows that Sp$(G)= \{\pm i t_\nu\}_{\nu=1}^{L/2},$ on the other hand from
(\ref{G}) one has Sp$(G)=\{i\tan(\theta_\mu/2)\}_{\mu=1}^L$ where $\{e^{i\theta_\mu}\}$ is the spectrum of the unitary $T.$
As stressed above we can  assume
$T\in SO_L(\mathbb{R})$  and then $T=e^K$ where $K\in o_L(\mathbb{R})$ i.e., it is skew-symmetric. One can now apply the decomposition of \cite{schl}
to $K$ i.e.,  $K_D= U K U^T=\oplus_{\nu=1}^{L/2} i\theta_\nu \sigma^y_{(\nu)},$ ( $\theta_\nu$'s can vanish)
and obtain Sp$(T)=\{e^{\pm i \theta_\nu}\}_{\nu=1}^{L/2}.$
It follows that $t_\nu=\tan(\theta_\nu/2)$ and from this (\ref{Psi_D}) is obtained.
$\hfill\Box$

To illustrate the formalism now we shall analyze the basic and elementary case of a pair of fermionic modes. 

{\em Example 1.}
Two fermionic modes:  $A=\epsilon \sigma^z, \, B=i\Delta \sigma^y.$
Then $H= \epsilon (c_1^\dagger c_1 - c_2^\dagger c_2) +\Delta (c_1^\dagger c_2^\dagger+\rm {h.c.})$,
from which $Z= \epsilon \sigma^z -i\Delta \sigma^y,\, ZZ^\dagger= (\epsilon^2+\Delta^2)\openone+2\Delta\epsilon \sigma^x$
$\Lambda={\rm{diag}}(|\epsilon -\Delta|, \,|\epsilon +\Delta|).$ By explicit computation of $T=|Z|^{-1}Z$ one finds that $T$ is  either
$\sigma^x$ or $i\sigma^y$.
Of course this model is trivially soluble with eigenvectors
$\{|10\rangle, |01\rangle, |00\rangle+|11\rangle, |00\rangle-|11\rangle \}$ and eigenvalues $\{\epsilon, -\epsilon,\Delta, -\Delta\}.$
The zero-eigenvalue lines $\epsilon =\pm\Delta$  splits the $\epsilon-\Delta$ plane in four regions, in each of which the ground-state is given by one 
of the states above.  
By considering instead the two fermion system with $Z=\sigma_x+\mu \openone-i\Delta \sigma^y$ (see the complete graph described below)
one finds that $T=\sigma_x$ ($ T=e^{-i\theta \sigma^y},\,\theta=\tan^{-1}\Delta/\mu$) for $\mu^2+\Delta^2\le 1$ ($\mu^2+\Delta^2>  1).$
The unit circle dividing the two regions is the set of points where $Z$ is singular.

{\em Example 2.} Two fermionic modes:  $A=\epsilon \openone, \, B=i\Delta \sigma^y.$
Then $H= \epsilon (c_1^\dagger c_1 + c_2^\dagger c_2) +\Delta (c_1^\dagger c_2^\dagger+\rm {h.c.})$,
from which $Z= \epsilon \openone -i\Delta \sigma^y,\, ZZ^\dagger= (\epsilon^2+\Delta^2)\openone,\,
\Phi=\openone,\,\Lambda= \sqrt{\epsilon^2+\Delta^2} \openone,\,T= \exp(-i\theta \sigma^y)$ where $\theta:=\tan^{-1}(\Delta/\epsilon).$
Therefore
$G=(e^{-i\theta \sigma^y}-\openone)(e^{-i\theta \sigma^y}+\openone)^{-1}=-i\tan(\frac{\theta}{2} \sigma^y)$.
At variance with the former case now we have zero single-particle eigenvalues
only in the trivial case $\epsilon=\Delta=0$ and $T(\theta)$ is a smooth family
of commuting matrices. This example can be readily extended to $L=2M$ fermionic
modes by considering  $A=\oplus_{\nu=1}^M  \epsilon_\nu \openone_\nu,\, B= i
\oplus_{\nu=1}^M \Delta_\nu \sigma^y_{(\nu)}.$
Then one finds $T=\oplus_{\nu=1}^M \exp(-i\theta_\nu \sigma^y_{(\nu)}),\,\theta_\nu=\tan^{-1}(\Delta_\nu/\epsilon_\nu),$



{\em The fidelity}--
We give now an explicit evaluation of the 
of the fidelity  (\ref{over}). 
In order to do that we first recall the following result from Ref.~\cite{perelemov}
\begin{equation}
\langle \Psi_{\tilde{Z}},\Psi_{{Z}}\rangle=\frac{ \det(\openone + G^\dagger \tilde G)^{1/2} }
{\det(\openone + G^\dagger G)^{1/4}\det(\openone + \tilde{G}^\dagger \tilde{G})^{1/4}} \ .
\label{overlap}
\end{equation}
%

{\em Proposition 1}
\begin{equation}
{\cal F}(Z, \tilde{Z})= \frac{1}{2^{L/2}}|\det( T +\tilde{T})|^{1/2}
=\prod_{\nu=1}^{L/2} |\cos(\Theta_\nu/2)|.
\label{overlap-marco}
\end{equation}
where the second equality holds for $T^{-1}\tilde{T}\in SO_L(\mathbb{R})$ and 
Sp$(T^{-1}\tilde{T})= \{e^{\pm i\Theta_\nu}\}_{\nu=1}^{L/2}.$
When $\det(T^{-1}\tilde{T})=-1$
one has ${\cal F}(Z, \tilde{Z})=0.$

{\em Proof.} 
To prove the first equality  it is a straightforward calculation from Eqs. (\ref{overlap}), (\ref{G}) and  $|\det T|=|\det\tilde{T}|=1$.
Of course if $-1\in{\rm{Sp}}(T^{-1}\tilde{T})$  then ${\cal F}(Z, \tilde{Z})=0$ but this happens ncessarily when  $\det(T^{-1}\tilde{T})=-1.$ 
If $T^{-1}\tilde{T}\in SO_L(\mathbb{R})$ one has
Sp$(T^{-1}\tilde{T})=\{e^{i\Theta_\mu}\}_{\mu=1}^L=\{e^{\pm i\Theta_\nu}\}_{\nu=1}^{L/2}$ (first equality from unitarity, second one
from reality and speciality) thus
one finds  ${\cal F}(Z, \tilde{Z})=\prod_{\mu=1}^L \sqrt{(1+e^{i\Theta_\mu})/{2}}=\prod_{\nu=1}^{L/2} |\cos(\Theta_\nu/2)|$.
$\hfill\Box$

When
 $[T,\,\tilde{T}]=0$ one has $\Theta_\mu=0, (\mu=1,\ldots,L-2M),\,\Theta_\nu= \pm(\theta_\nu-\tilde{\theta}_\nu),
\,(\nu=1,\ldots,M);$ therefore one gets the following result extending the analogous one obtained for the XY model \cite{za-pa}:
suppose now that $Z$ and $\tilde{Z}$ are such that $[T,\,\tilde{T}]=0$, then
${\cal F}(Z, \tilde{Z})
=\prod_{\nu=1}^{M}\cos(\frac{\theta_\nu -\tilde{\theta}_\nu}{2})$.
For example the Hamiltonian of {\em Ex. 1}, for all possible choices of $\{\epsilon\}_\nu $ and $\{\Delta_\nu\}_\nu,$ 
gives rise to commuting $T$'s .

To summarize: the conceptual path to find the fidelity between two different ground states of (\ref{ham}) is described by the following  chain of maps
$Z, \tilde{Z}\in M_L(\mathbb{R})\longrightarrow T, \tilde{T} \in  O_L(\mathbb{R})\longrightarrow {\rm{Sp}}(T^{-1}\tilde{T})\subset S^1.
$
The first arrow denotes the passage from the matrices $Z, \tilde{Z}$ to the orthogonal matrices  $T$ and $\tilde{T}$ i.e., their
polar parts, the second denotes the diagonalization of $T^{-1}\tilde{T}.$  
While the positive part of $Z$ i.e., $|Z|,$ contains the single-particle energy spectrum, in its polar part $T$ are  encoded all the data
to defining the many-body ground state (\ref{Psi_D}). In the points where $Z$ becomes singular i.e., zeroes in the single-particle spectrum,
the polar part is not uniquely defined:
by crossing these points one can have a non smooth  change of the $T,$ which in turn results in a non smooth change of the ground-state structure. 
If singularity of $Z$ is achieved for finite (infinite) size $L$ one has a first (higher) order QPT.
This kind of singular behaviour is reflected in 
Eq.  (\ref{overlap-marco}) which clearly shows that an {\em abrupt decrease of the fidelity, moving 
from $Z$ to the neighbouring $\tilde{Z},$ can be caused by the appearence of an eigenvalue of $T^{-1}\tilde{T}$ with a `large' negative part;
the closer this latter gets to $-1$ the smaller the  fidelity.}

An instance of this phenomenon is of course provided by the discontinuous ``first order" QPTs of {\em Example 1} 
in that  $\det(T^{-1}\tilde{T})=-1$ when $T$ and $\tilde T$ are in different phases.
A less trivial one can be obtained from the multi mode  case in {\em Ex. 2}. Suppose that, for some $\nu_0,$ one finds e.g., in the thermodynamical limit,
that $\Delta_{\nu_0}\mapsto 0^+$ and then  $\theta_{\nu_0}=\lim_{\Delta_{\nu_0}\to 0}\cos^{-1}(\epsilon_{\nu_0}/(\Delta_{\nu_0}^2+ \epsilon_{\nu_0}^2)^{1/2})
=(\pi/2)(1-\epsilon_{\nu_0}/|\epsilon_{\nu_0}|).$ If $\epsilon_{\nu_0}$ can be driven (in the same limit)
throughout zero one observes a jump of $\theta_{\nu_0}$ with amplitude $\pi;$ this in turn implies $-1$ is in the spectrum of $T^{-1}\tilde{T}$
[$T=T(\epsilon_{\nu_0}=0^+)$ and  $\tilde{T}=\tilde{T}(\epsilon_{\nu_0}=0^-)].$  This  is the mechanism responsible for the  fidelity drop in the $XY$
model observed in the paramagnetic-ferromagnetic QPT \cite{para-ferro}. The important aspect is that {\em the drop of (\ref{overlap}) corresponds here 
at $\epsilon_{\nu_0}^2+\Delta_{\nu_0}^2=0$ i.e., the vanishing of one of the single-particle energies, or, equivalently, a gaplessness in the many-body
energy spectrum.}
This kind of  correspondence between the  fidelity behaviour and zeroes in single particle spectrum
can be further unveiled  by using the following "perturbative" result

{\em Proposition 2}
If $T(\lambda)=e^{K(\lambda)}, K\in o_L(\mathbb{R})$ and $Z=Z(\lambda), \tilde{Z}=Z(\lambda+ \delta\lambda)$
one has ${\cal F}(\tilde{Z},{Z})=e^{-S(Z)}$ where
\begin{equation}
{S}({Z})=-\frac{1}{16} {\rm{Tr}} (K^\prime \delta\lambda)^2+ O(\delta\lambda^3)
\label{expand}
\end{equation}
where $ K^\prime:={\partial K}/{\partial\lambda}.$

{\em Proof.}
From Eq. (\ref{overlap-marco}), using $\det A=\exp(\rm{Tr}\log A),$ one has
${\cal F}(\tilde{Z},Z) = \exp\{ (1/2) \rm{Tr} \ log[(1+ T^{-1}\tilde{T})/2]\},$ by 
using the Taylor expansion of the logarithm one obtains  $e^{S(\lambda)},\, S(\lambda)=\sum_{n=0}^\infty  L_n$ where
 $L_n=\frac{(-1)^{n-1}}{n 2^{n+1}} {\rm{Tr}}(T^{-1}\delta T)^n$
Now one can write  $T(\lambda)=e^{K(\lambda)}, K\in o_L(\mathbb{R})$ then $\delta T= \tilde{T}-T= K^\prime T \delta\lambda + (1/2)[ K^{\prime\prime} T + (K^\prime)^2 T]
\delta\lambda^2  +O(\delta\lambda^3)$
(prime denotes derivation).
From unitarity one has that $K^\prime, K^{\prime\prime}$ are  traceless, hence $L_1\approx 1/8 {\rm{Tr}} (K^\prime \delta\lambda)^2 $ and 
$L_2 \approx -1/16 (K^\prime \delta\lambda)^2.$
From here using the above expansion one obtains Eq.~(\ref{expand}).
$\hfill\Box$

To illustrate this result let us consider the system in {\em Example 2} i.e., $K(\lambda)=i \sum_\nu \theta_\nu(\lambda) \sigma^y_{(\nu)}$ with
$\epsilon_\nu$ and $\Delta_\nu$ functions of the parameter $\lambda.$
  In this case, by taking the derivative of 
$\theta_\nu=\tan^{-1}(\Delta_\nu/\epsilon_\nu)$ one gets $\rm{Tr}(K^\prime)^2=-2 \sum_\nu (\theta^\prime_\nu)^2$ and hence
${\cal F}(\tilde{Z},{Z}) =e^{-\delta\lambda^2 S_2 + o(\delta\lambda^3)}$,
where
$S_2= 1/8\sum_\nu (\frac{\epsilon_\nu^2}{\epsilon_\nu^2+\Delta_\nu^2} D^\nu_Z)^2 
$
with $D^\nu_Z= \epsilon_\nu^{-1} \Delta_\nu' - \epsilon_\nu^{-2}\Delta_\nu \epsilon_\nu'$. 
This equation   suggests  that {\em if some of the $|Z|$ eigenvalues $\sqrt{\epsilon_\nu^2+\Delta_\nu^2}$ is zero, or asymptically 
vanishing for $L\rightarrow\infty,$ then the 
function $S_2$ should have a sharp increase and accordingly the  fidelity  a sharp decrease}. 



{\em Complete graph}--
In the remaining of the paper we will provide a GS fidelity study for the
fermionic system (\ref{ham}) with an underlying topology of a {\em complete}
graph. A detailed analysis of this diagram will be reported in a separated publication \cite{cozzini}.
More precisely, now we consider the Hamiltonian (\ref{ham}) defined by the
following matrix data
%
$A(\mu)_{ij} = 1+(\mu-1)\delta_{ij} \,,\;
B(\gamma)_{ij} = \gamma\,{\rm{sign}}(j-i)
$
%
where $(i,j=1,\ldots,L).$ 
We analyze now some simple cases in the $(\mu,\gamma)$ plane.

i) For $\gamma=0$ the corresponding number-conserving single-particle
Hamiltonian is readily diagonalized: ${\rm Sp}\,A(\mu)=\{L+\mu-1,\mu-1\},$
where the first (second) eigenvalue has degeneracy one ($L-1$). We see that
$\mu=1$ corresponds to the vanishing of the lowest eigenvalue (for any $L$).
ii) For $(\mu=0,\gamma=1)$ the matrices $Z$ and $Z^\dagger=Z^T$ become lower and upper
triangular respectively. By explict computation one finds
$(ZZ^\dagger)_{ij}=4\min(L-i, L-j)$, which has the last column and row identically
vanishing. Accordingly $0\in\text{Sp}\,|Z(0,1)|\;\forall\;{L}$.
%
%
iii) Changing the sign of $\gamma$ simply corresponds to
transforming $Z$ into $Z^\dagger$. Since $\text{Sp}(ZZ^\dagger)=\text{Sp}(Z^\dagger Z)$, the single
particle spectrum is not affected by the transformation. Furthermore, since
$T(-\gamma)=T^\dagger(\gamma)$ and $\det{T}=\det{T^\dagger}$, the overlap behaviour is
symmetric with respect to the $\gamma=0$ axis.

\begin{figure}[h]
\includegraphics[width=7cm,height=6.6cm]{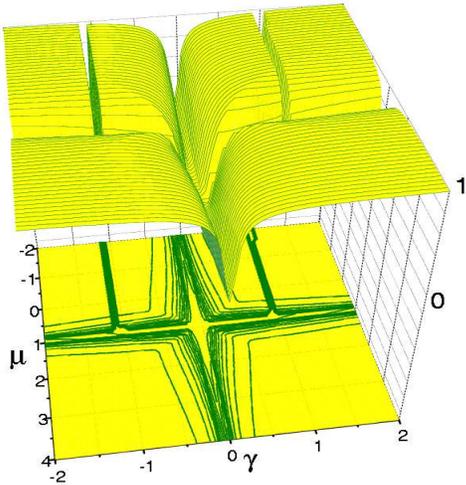}
\caption{\label{fig:1}%
GS fidelity in the complete graph for $L=400$.
The plotted function is
$\mathcal{F}_{\text{min}}=\min[\mathcal{F}(Z,\tilde{Z}_{\delta\mu}),
\mathcal{F}(Z,\tilde{Z}_{\delta\gamma})]$ (see text),
with $\delta\mu=\delta\gamma=0.1$.
Note the symmetry with respect to the $\gamma=0$ axis.}
\end{figure}

\begin{figure}[b]
\includegraphics[width=4cm]{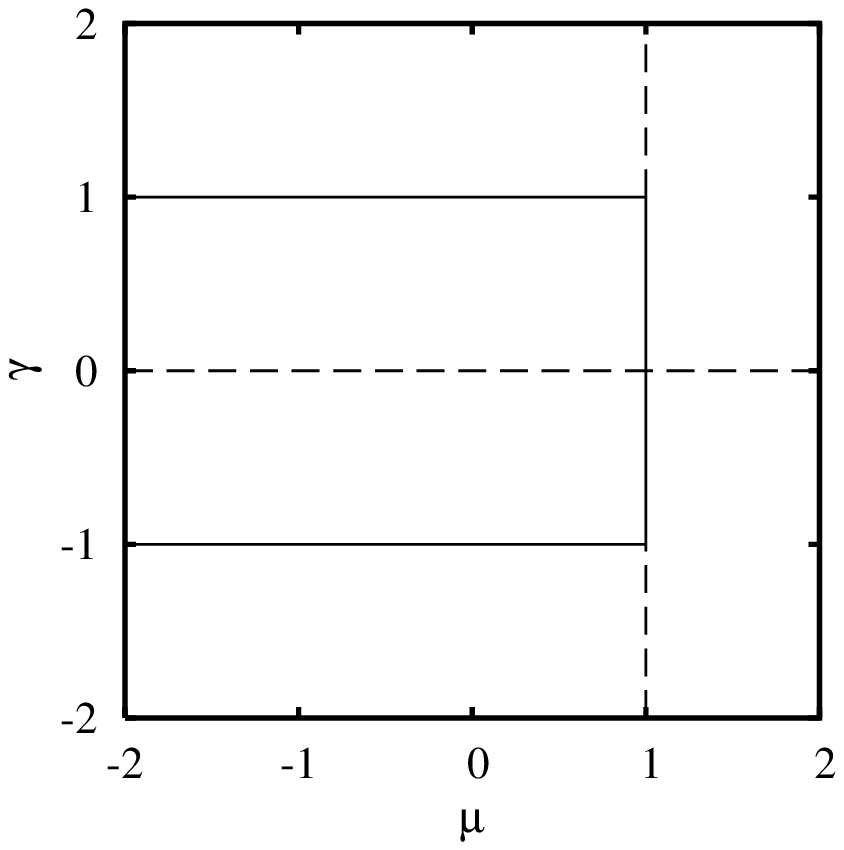}
\includegraphics[width=4cm]{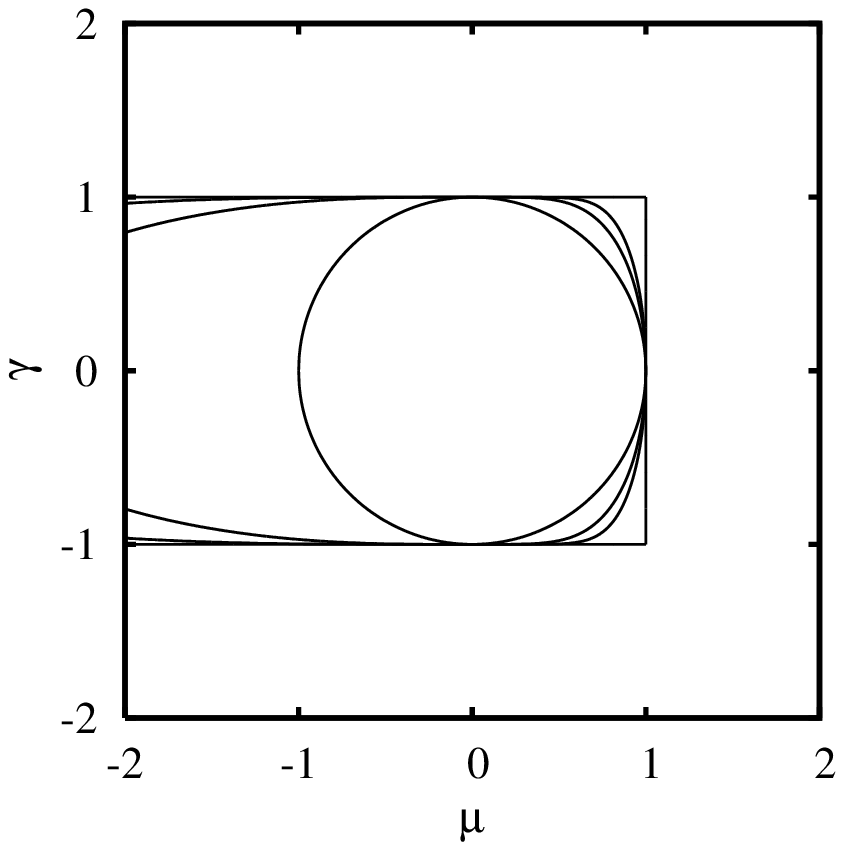}
\caption{\label{fig:2}%
Phase diagram in the $\mu$-$\gamma$ plane for $L$ even.
Solid and dashed lines correspond to first and higher order QPTs
respectively.
Left panel: thermodynamic limit $L\to\infty$.
Right panel: phase boundaries of the first order QPT for $L=2,4,6,\infty$.}
\end{figure}

The GS fidelity in the $\mu$-$\gamma$ plane is reported in Fig.~\ref{fig:1} for
$L=400$. By using
$\mathcal{F}_{\text{min}}=\min[\mathcal{F}(Z,\tilde{Z}_{\delta\mu}),
\mathcal{F}(Z,\tilde{Z}_{\delta\gamma})]$, where
$\tilde{Z}_{\delta\mu}=Z(\mu+\delta\mu,\gamma)$ and
$\tilde{Z}_{\delta\gamma}=Z(\mu,\gamma+\delta\gamma)$, i.e., by plotting the
minimum of the fidelity with respect to variations in both directions, one gets
a clear diagram of the `degree of orthogonality' of neighbouring ground states
in the parameter space.
The drops in the GS overlap identify all the QPTs of the system. In general,
the actual value of the drop at a given point in the $\mu$-$\gamma$ plane
depends both on the chosen variations $\delta\mu$, $\delta\gamma$, and on the
system size $L$.
For even values of $L$ and in the thermodynamic limit, the phase diagram is
reported in the left panel of Fig.~\ref{fig:2}.
The boundary of the region $\mu<1$, $|\gamma|<1$ corresponds to a first order
QPT. Indeed, while the ground state inside this region contains an odd number
of fermions ($\det{T}=-1$), it becomes even outside ($\det{T}=1$). Such a QPT
is simply interpreted in terms of a level crossing in the energy spectrum.
The phase diagram of this first order QPT can be clearly calculated also for
finite values of $L$. 
 In the right panel of Fig.~\ref{fig:2} the
cases $L=2,4,6,\infty$ are shown.
The line $\mu=1$, for $|\gamma|>1$, marks instead a second order QPT,
accompanied by the vanishing of the energy gap in the thermodynamic limit. Here
the fidelity vanishes only in the thermodynamic limit 
\cite{ortho-cata}. 

Finally, we observe a similar drop in the fidelity also at $\gamma=0$, for any
value of $\mu$. Here the model is exactly solvable and, as mentioned above, the
energy gap in the thermodynamic limit is given by $\mu-1$.
In spite of the gapfulness of the spectrum, we identify the fidelity drop with
a higher order QPT.
Indeed, the line $\gamma=0$ shows the same finite-size scaling
properties as the transition for $\mu=1$, $|\gamma|>1$.

{\em Conclusions}-- In this paper we analyzed
general systems of free-fermions.
We have showed that the information about the zero-temperature phase diagram of these linear system  is encoded in the polar decomposition 
of the  $L\times L$ matrix of their coupling constants; the analyticity properties of this decomposition
dictate those of the ground state. We have given an explict expression for the fidelity between different ground states in terms
of the unitary parts of the corresponding polar decompositions and analyzed the ground state phase diagram in terms of  it.
In particular we showed a connection between the gaplessness of the many-body
energy spectrum and the fidelity drops.
Finally, to exemplify the general results of the paper,  we have 
presented a preliminary study of the ground-state diagram of a free-fermi system over a totally connected graph.
What appears to be rather  remarkable in this fidelity approach lies in its universality and purely (Hilbert space) geometrical
nature
: no a priori understanding of the nature of the quantum phases
separated by critical points is required.




\begin{thebibliography}{99}
\bibitem{losch} The quantity (\ref{over}) represent a generalization of the Loschmidt echo analyzed, in relation to QPTs,
in Ref. \cite{zanardi-china} and in \cite{cuc}. To see this notice first that the bilinears $\{ c_i^\dagger c_j, c_i^\dagger c_j^\dagger, \ldots\}$
of fermionic operators  close into a Lie-algebra $so(2L, \mathbb{R})$ \cite{perelemov}, this  implies that $\exp(i H_W t) H_Z \exp(-i H_W t)= H_{Z(t)}.$
and thus that $e^{i H_W t}|\Psi_Z\rangle=|\Psi_{Z(t)}\rangle.$ Moreover
${\cal F}(0,t)=|\langle\Psi_{Z(0)}, \Psi_{Z(t)}\rangle|=|\langle\Psi_{Z(0)},
e^{i H_W t} \Psi_{Z(0)}\rangle|.$ When $W=Z+\delta Z$  the latter expression gives the   Loschmidt echo.
\bibitem{para-ferro} In that case $\epsilon_{\nu_0}=\cos(2\pi/L)-\lambda,\,\Delta_{\nu_0}=\gamma \sin(2\pi/L)$
where $\lambda$ and $\gamma$ are the external magnetic field and anisotropy parameter respectively.
\bibitem{ortho-cata}
Unless the ground state is constant in a given phase,
the fidelity function always vanishes in the thermodynamic limit.
However, the rate of decrease in the overlap is much faster in the proximity of
QPTs.
\end{thebibliography}
\end{document}